\documentclass[11pt,english]{article}
\usepackage[T1]{fontenc}
\usepackage[latin9]{inputenc}
\usepackage{geometry}
\usepackage{xcolor}
\usepackage{textcomp}
\usepackage{amssymb}
\usepackage{graphicx}
\usepackage{setspace}
\usepackage{esint}
\makeatletter

\providecommand{\tabularnewline}{\\}

\newcommand{\lyxaddress}[1]{
	\par {\raggedright #1
	\vspace{1.4em}
	\noindent\par}
}

\makeatother

\usepackage{babel}
\begin{document}
\title{E\textsuperscript{2} distribution and statistical regularity in polygonal
planar tessellations}
\author{Ran Li\textsuperscript{1}, Consuelo Ibar\textsuperscript{2}, Zhenru
Zhou\textsuperscript{2}, Seyedsajad Moazzeni\textsuperscript{1},
\\Kenneth D. Irvine\textsuperscript{2}\thanks{Corresponding email: irvine@waksman.rutgers.edu},
Liping Liu\textsuperscript{1,3}\thanks{Corresponding email: liu.liping@rutgers.edu}
, and Hao Lin\textsuperscript{1}\thanks{Corresponding email: hlin@soe.rutgers.edu}}
\maketitle

\lyxaddress{1. Department of Mechanical and Aerospace Engineering, Rutgers, The
State University of New Jersey \\ 2. Waksman Institute and Department
of Molecular Biology and Biochemistry, Rutgers, The State University
of New Jersey\\3. Department of Mathematics, Rutgers, The State University
of New Jersey}
\begin{abstract}
From solar supergranulation to salt flat in Bolivia, from veins on
leaves to cells on Drosophila wing discs, polygon-based networks exhibit
great complexities, yet similarities persist and statistical distributions
can be remarkably consistent. Based on analysis of 99 polygonal tessellations
of a wide variety of physical origins, this work demonstrates the
ubiquity of an exponential distribution in the squared norm of the
deformation tensor, E\textsuperscript{2}, which directly leads to
the ubiquitous presence of Gamma distributions in polygon aspect ratio.
The E\textsuperscript{2} distribution in turn arises as a $\chi^{2}$-distribution,
and an analytical framework is developed to compute its statistics.
E\textsuperscript{2} is closely related to many energy forms, and
its Boltzmann-like feature allows the definition of a pseudo-temperature.
Together with normality in other key variables such as vertex displacement,
this work reveals regularities universally present in all systems
alike.
\end{abstract}
Polygonal networks are one of nature's favorite ways of organizing
the multitude - from supergranulation on the solar surface \cite{Hirzberger2008}
to cracked dry earth \cite{Crack8}; from ice wedges in northern Canada
\cite{Ice4} to the scenic Salar de Uyuni in Bolivia \cite{SaltFlat1};
and from veins on leaves \cite{Leaf2} to cells on Drosophila wing
discs \cite{Pan2016} (Fig. \ref{fig:Schematic}). These systems are
driven by distinctive physical mechanisms, yet they share common features.
Individual constituents, namely, ``cells'' appear to ``randomize''
into statistical distributions, and only interact with their immediate
neighbors. On the collective level, especially in the dynamic and
active systems, rich phenomena are observed, including unjamming and
jamming, fluid-to-solid phase transition, and flow and migration\textcolor{black}{{}
\cite{Bi2015,Farhadifar2007,Garcia2015,KimHilgenfeldtSM2015,LiuNagle1998,Nnetu2013,Park2015,Sadati2013,Trappe2001}.}

Despite the complexity and variabilities involved in these phenomena,
similarity patterns emerge. One particularly interesting instance
is provided recently by Atia \emph{et al. }\cite{Atia2018}. Within
the context of confluent biological tissue and based on extensive
experiments both \emph{in vitro} and \emph{in vivo}, the authors found
that data on cell aspect ratio collapse and follow a normalized Gamma
distribution, implying a universal principle underlying the geometric
configuration and pertinent processes.

What is the basis of this universality? Does it carry beyond the biological
context? This work composes of a two-part discovery to answer these
questions. In the first, we analyze a total of 99 data sets in 8 groups
that include convection patterns (solar supergranulation), landforms
(salt flats, on Mars, and in or near the Arctic), cracked dry earth,
and biological patterns (veins on leaves, cells on Drosophila wing
discs, and plated MDCK cells). (Fig. \ref{fig:Schematic} and Table
\ref{tab:R2}.) We demonstrate that the Gamma distribution in polygon
aspect ratio derives from an exponential distribution in the squared
strain tensor norm, E\textsuperscript{2}, based on which we develop
a unifying solution for the former. Both distributions persist in
all data examined. In the second, we tackle the origin of the E\textsuperscript{2}
distribution, and demonstrate that it arises as a $\chi^{2}$-distribution
owing to asymptotic normality of vertex displacement and other key
variables. We present a theoretical framework to accurately compute
E\textsuperscript{2} from vertex statistics. Importantly, E\textsuperscript{2}
is closely related to various definitions of system energy including
all of the bulk-, perimeter-, and moment-based (known as the Quantizer)
forms \cite{Bi2015,Du2005Gersho,HainKlattTurk2020,KimHilgenfeldtSM2015,KlattNatCom2019,Manning_2010}.
The Boltzmann-like feature of E\textsuperscript{2} enables the definition
of a pseudo-temperature that can be considered as a consistent quantifier
\cite{DBB1985,Edwards1998,Edwards1989,McNamara2009}. The exponential
and normal distributions we reveal in this work are hidden regularities
that transcend the specific physical systems, and are embedded universal
features of random polygonal tessellations.

\subsection*{E\textsuperscript{2} and its relationship with aspect ratio}

\begin{figure}
\center\hspace{-.5cm}\includegraphics[clip,width=1\linewidth]{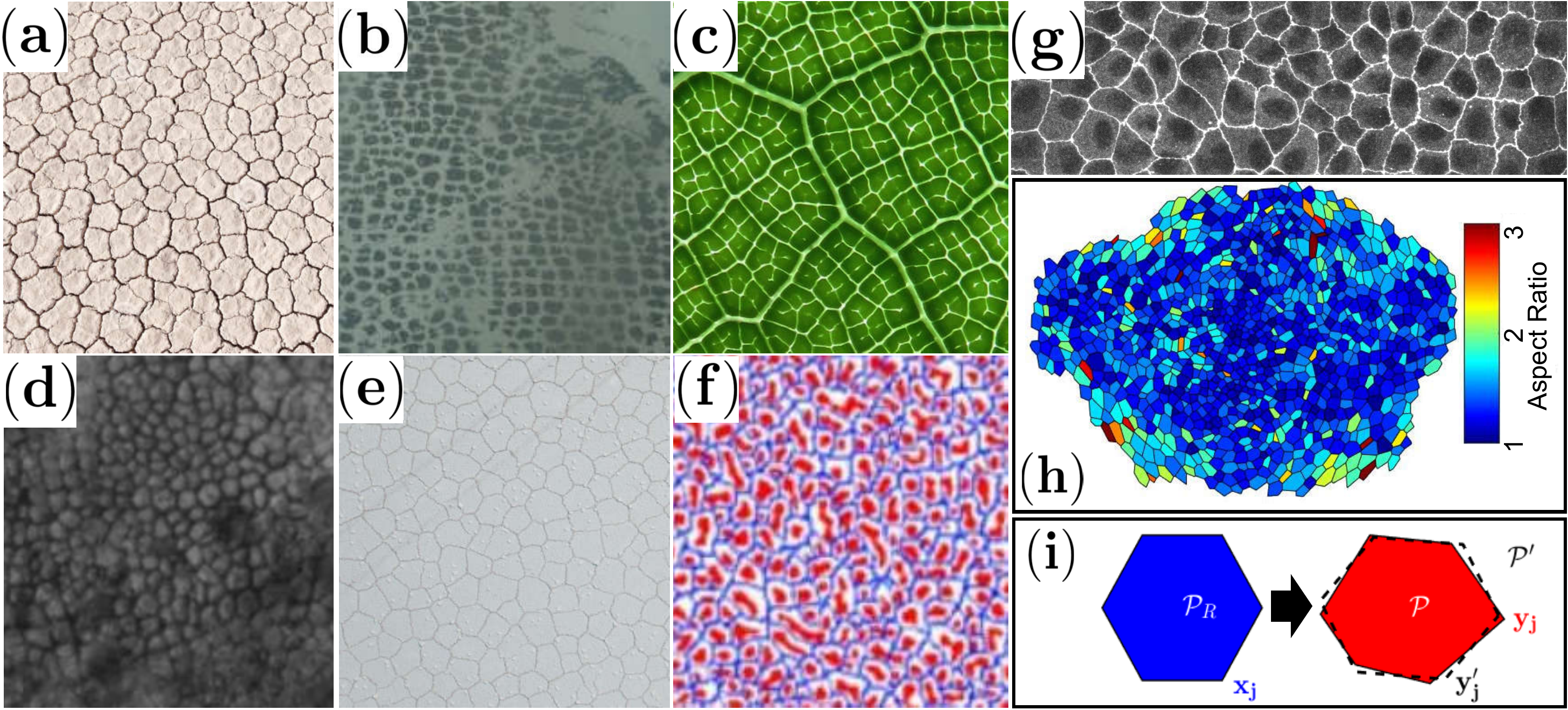}

\caption{(a)-(h) Examples of randomized polygonal networks in nature. (a) Land
cracks due to desiccation \cite{Crack8}; (b) Ice wedges from northern
Canada \cite{Ice4}; (c) Veins on a \emph{Ficus lyrata} leaf \cite{Leaf2};
(d) Desiccation pattern of an ancient lake on Mars \cite{ElMaarry2010};
(e) A snapshot of Salar de Uyuni in Bolivia, world's largest salt
flat \cite{SaltFlat1}; (f) Supergranulation on solar surface \cite{Hirzberger2008};
(g) \textcolor{black}{Plated MDCK cells (this work).}\textcolor{orange}{{}
}(h) A processed image of a developing Drosophila wing disc (this
work) with the aspect ratio of cells color-mapped. (i) A regular hexagon
$\mathcal{P}_{R}$ (blue, with vertices ${\bf x}_{j}$), a deformed
hexagon $\mathcal{P}$ (red, with vertices ${\bf y}_{j}$), and a
uniform deformation as mean-field approximation $\mathcal{P}'$ (black
dashed outline, with vertices ${\bf y}_{j}'$). \label{fig:Schematic}}
\end{figure}

In this first part of this work, we define E\textsuperscript{2} and
analytically establish its relationship with the polygon aspect ratio.
We demonstrate that a $k$-Gamma distribution in the latter is derivable
from an exponential distribution in the former. We begin by defining
the mean-field deformation tensor, ${\bf E}$. An exemplary processed
image of a Drosophila wing disc 120h after egg laying (AEL) is shown
in Figure \ref{fig:Schematic}h, where the color scale indicates magnitude
of the cell aspect ratio (defined in Methods). We choose as our reference
frame a \emph{regular} $n$-polygon centered at the origin, with vertices
\begin{equation}
{\bf x}_{j}=r_{0}{\bf e}_{j},\quad{\bf e}_{j}=[\cos(j2\pi/n),\:\sin(j2\pi/n)],\quad j=1,...,n.\label{eq:x0}
\end{equation}
This polygon is denoted by $\mathcal{P}_{R}$, and $r_{0}$ is a scaling
factor. Figure \ref{fig:Schematic}k uses $n=6$, a hexagon as an
illustrative example. We consequently regard any $n$-sided polygon
$\mathcal{P}$ with vertices ${\bf y}_{j}$ as a ``deformation''
from $\mathcal{P}_{R}$,
\begin{equation}
{\bf y}_{j}={\bf x}_{j}+{\bf u}_{j},\quad j=1,...n.\label{eq:uj}
\end{equation}
Note that this deformation is to be understood as a morphological
deviation from the reference $\mathcal{P}_{R}$ , rather than an actual
physical deformation, although the latter is a possibility. We also
require that the centroid of $\mathcal{P}$ is aligned with that of
$\mathcal{P}_{R}$. The deformation is in general non-uniform, in
the sense that for $n\ge3$, a single deformation tensor of ${\bf F}\in\mathbb{R}^{2\times2}$
cannot be identified by $\mathbf{y}_{j}=\mathbf{Fx}_{j}$ for all
of $j=1,...,n$. Nevertheless we \emph{can} introduce an approximation,
namely,
\begin{equation}
\mathbf{y}_{j}\approx{\bf y}_{j}'=\mathbf{Fx}_{j},\quad j=1,...,n.\label{eq:yj}
\end{equation}
This approximation is analogous to a Taylor expansion in which only
the leading order term is retained. The use of a uniform deformation
to approximate the local and non-uniform deformation field is effectively
coarse-graining, reducing the degree of freedom from $2n$ to $4$
and suppressing the fluctuations. This idea is illustrated in Figure
\ref{fig:Schematic}i (right), where $\mathcal{P}'$ is the approximate
and uniformly deformed polygon. From ${\bf F}$ we define the usual
strain tensor and its squared Frobenius norm,
\begin{equation}
\mathbf{E}=({\bf F}^{\mathrm{T}}{\bf F})^{\frac{1}{2}}-{\bf I}\approx\frac{1}{2}[(\mathbf{F}-\mathbf{I})+(\mathbf{F}-\mathbf{I})^{\mathrm{T}}],\quad|{\bf E}|^{2}={\rm Tr}({\bf E}^{{\rm T}}{\bf E}),\label{eq:Strain}
\end{equation}
where ${\bf I}$ is the identity tensor, and the approximation in
the first equation is valid in the small-to-moderate deformation regime.
In this work we use $|{\bf E}|^{2}$ (mathematical representation)
and E\textsuperscript{2} (terminology) interchangeably. The restriction
to isochoric (area-conserved) deformation requires that $\det{\bf F}=1$
and equivalently (to leading order), ${\rm Tr}{\bf E}=0$, which is
satisfied by choosing $r_{0}$ in (\ref{eq:x0}). Consequently, ${\bf E}$
has a small degree of freedom of $2$, which eventually leads to its
strong regularity that we demonstrate later.

We pursue an analytical expression for ${\bf E}$ via minimizing the
difference between ${\bf y}_{j}$'s and ${\bf y}_{j}'$'s in (\ref{eq:yj}),
from which we obtain (SI) 
\begin{equation}
{\bf E}=\frac{1}{n}\sum_{j=1}^{n}[{\bf v}_{j}\otimes{\bf e}_{j}+{\bf e}_{j}\otimes{\bf v}_{j}-({\bf e}_{j}\cdot{\bf v}_{j}){\bf I}],\label{eq:cE}
\end{equation}
\begin{equation}
|{\bf E}|^{2}=\sum_{i=1}^{n}\sum_{j=1}^{n}{\bf v}_{i}\cdot{\bf C}_{ij}{\bf v}_{j},\quad{\bf C}_{ij}=\frac{2}{n^{2}}[({\bf e}_{i}\cdot{\bf e}_{j}){\bf I}+{\bf e}_{j}\otimes{\bf e}_{i}-{\bf e}_{i}\otimes{\bf e}_{j}].\label{eq:E2}
\end{equation}
Here 
\begin{equation}
{\bf v}_{j}={\bf u}_{j}/r_{0},\quad r_{0}=\frac{1}{n}\sum_{j=1}^{n}{\bf y}_{j}\cdot{\bf e}_{j}.\label{eq:vi}
\end{equation}
In addition to satisfying the trace-free condition, the scaling also
ensures that ${\bf v}_{j}$'s are dimensionless. Further details on
the computation of $|{\bf E}|^{2}$ is deferred until later. For now
we show that if E\textsuperscript{2} follows an exponential distribution
(that shall be validated below both via data and analysis), namely,

\begin{equation}
\rho_{{\rm E}}(|{\bf E}|^{2})=\beta\exp(-\beta|{\bf E}|^{2}),\label{eq:Kinematic}
\end{equation}
then the aspect ratio follows a $k$-Gamma distribution such as shown
in \cite{Atia2018}. Here $\rho(\cdot)$ denotes a (normalized) probability
density function (PDF), and $\beta$ is similar to an inverse temperature
as the PDF is Boltzmann-like. The aspect ratio $a_{r}$ of the polygon
is calculated via the second area moments (Methods), and is related
to E\textsuperscript{2} by (SI)
\begin{equation}
a_{r}^{2}=(x+1)^{2}=g(|{\bf E}|^{2}),\quad g(t)=1+2t^{2}+4t+[(2t^{2}+4t+1)^{2}-1]^{1/2},\label{eq:xgE}
\end{equation}
where for convenience we define a shape factor $x$ in relation to
$a_{r}$ using $x=a_{r}-1$. Based on (\ref{eq:Kinematic}, \ref{eq:xgE}),
a transformation $\rho_{{\rm E}}(|{\bf E}|^{2}){\rm d}|{\bf E}|^{2}=\rho_{X}(x){\rm d}x$
leads to
\begin{equation}
\rho_{X}(x)=\rho_{{\rm E}}(|{\bf E}|^{2})\frac{{\rm d}|{\bf E}|^{2}}{{\rm d}x}=\beta\zeta'(x)\exp(-\beta\zeta(x)),\label{eq:rhox}
\end{equation}
where 
\[
\zeta(x)=g^{-1}((x+1)^{2}).
\]
Here we have considered the normalization condition. Equation (\ref{eq:rhox})
is our prediction of the distribution in the aspect ratio.

\subsection*{Data validate E\textsuperscript{2} and $a_{r}$ distributions}

\begin{figure}
\center\includegraphics[width=1\linewidth]{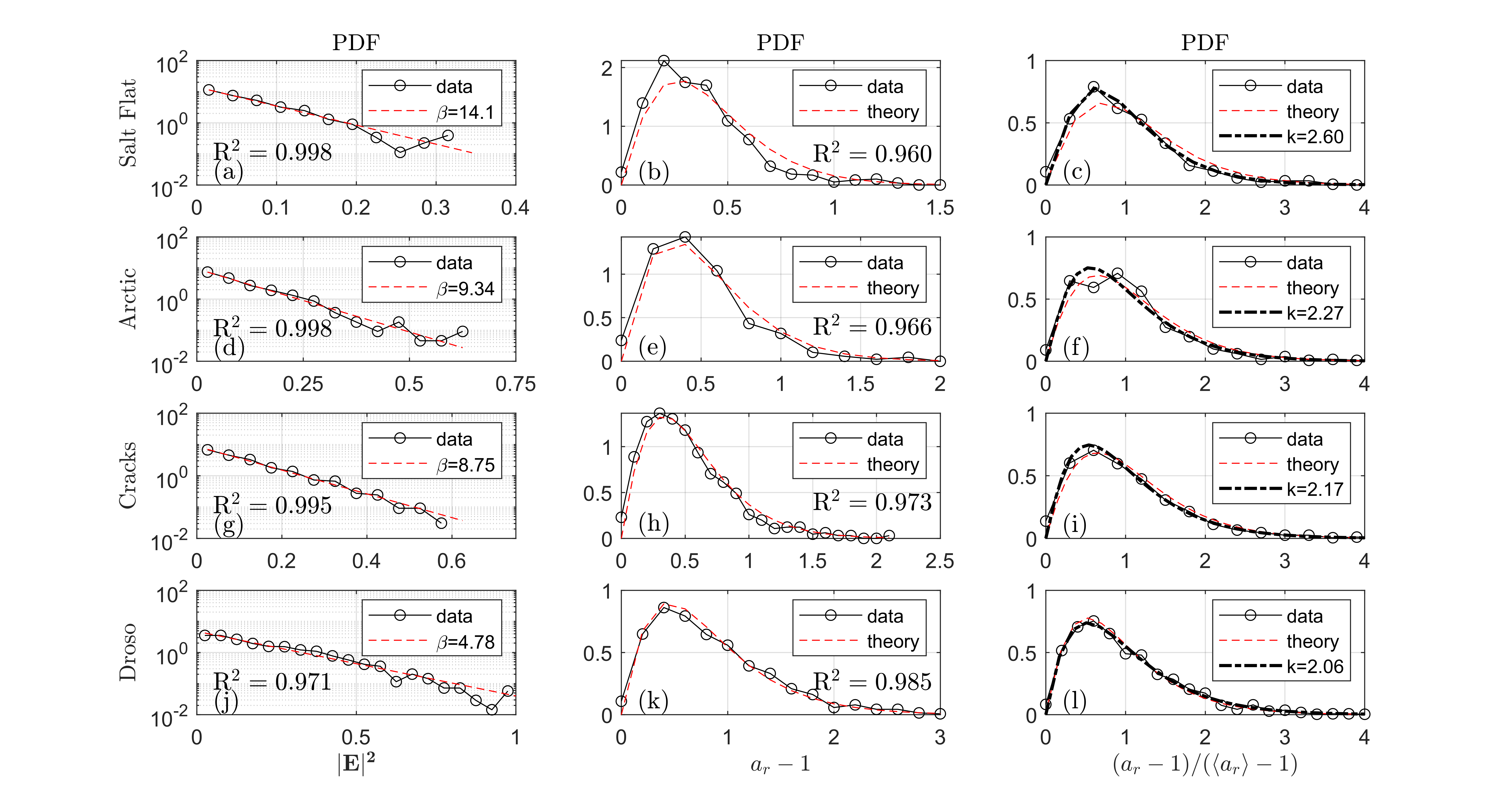}

\caption{Universality in strain and aspect ratio distributions. Left column
shows the PDFs of $|{\bf E}|^{2}$, fitted with an exponential form
$\exp(-\beta|{\bf E}|^{2})$ to extract $\beta$. This $\beta$ value
is used in Eq. (\ref{eq:rhox}) to generate the theoretical curves
in the middle column (dash), in comparison to the aspect ratio data
(symbols). The coefficients of determination, ${\rm R}^{2}$ are shown
within the panels. Right column: both data and theoretical curves
from the center column are normalized using the mean values, and fitted
with a $k$-Gamma function (\ref{eq:rhokG}) (thick dashed). A single
parameter $k$ is extracted and shown in the figure legends. Data
are from \cite{SaltFlat2},\textcolor{magenta}{{} }\cite{Ice6}, and
\cite{Crack10}, respectively, for the top 3 rows; and from this work
for Droso. \label{fig:Master}}
\end{figure}

Both distributions and their relationship are extensively validated
with a total of 99 data sets spanning 8 groups summarized in Table
\ref{tab:R2}; detailed descriptions, data sources, and method of
analysis are presented in the Methods section. Figure \ref{fig:Master}
uses 4 representative cases to demonstrate the agreement. (More are
shown in Fig. S2 in the SI.) The left column shows the PDFs of $|{\bf E}|^{2}$,
which are very well fitted by the exponential form, $\exp(-\beta|{\bf E}|^{2})$,
where $\beta$ is extracted as a fitting parameter. (See also Fig.
\ref{fig:UND}b where all $|{\bf E}|^{2}$ profiles are presented
and the value of $\beta$ is theoretically predicted.) 
\begin{table}
\center

\begin{tabular}{|c|c|c|c|c|}
\hline 
Type (abbreviation) & $M$ & $N$ & ${\rm R}^{2}$, $|{\bf E}|^{2}$ & ${\rm R}^{2}$, $a_{r}-1$\tabularnewline
\hline 
\hline 
Salt Flat of Uyuni (Salt Flat) & $7$ & $193-849$ & $0.994\pm0.0058$ & $0.939\pm0.0255$\tabularnewline
\hline 
Landforms on Mars (Mars) & $9$ & $219-5826$ & $0.986\pm0.0125$ & $0.935\pm0.0461$\tabularnewline
\hline 
Veins on Leaves (Leaves) & $6$ & $338-6050$ & $0.994\pm0.0047$ & $0.936\pm0.0328$\tabularnewline
\hline 
Landforms in the Arctic (Arctic) & $11$ & $104-1061$ & $0.982\pm0.0169$ & $0.902\pm0.0728$\tabularnewline
\hline 
Supergranulation on Solar Surface (Solar) & $9$ & $192-1645$ & $0.991\pm0.0075$ & $0.932\pm0.0463$\tabularnewline
\hline 
Cracked Dry Earth (Cracks) & $11$ & $298-1596$ & $0.992\pm0.0067$ & $0.943\pm0.0353$\tabularnewline
\hline 
Drosophila Wing Disc (Droso) & $42$ & $902-4205$ & $0.991\pm0.0083$ & $0.955\pm0.0335$\tabularnewline
\hline 
Plated MDCK Cells (MDCK) & $4$ & $1148-2283$ & $0.997\pm0.0012$ & $0.936\pm0.0157$\tabularnewline
\hline 
\end{tabular}

\caption{Summary of data for a total of $M_{{\rm tot}}=99$ tessellations.
Abbreviations are defined within parentheses and are used in figure
legends; $M$ is data sets in each type; $N$, the number of polygons
in each set (range provided). ${\rm R}^{2}$ for $|{\bf E}|^{2}$
indicates quality of fitting (e.g., in left column, Fig. \ref{fig:Master});
${\rm R}^{2}$ for $a_{r}-1$ indicates quality of agreement between
theory and data (e.g., in middle column, Fig. \ref{fig:Master}).
Data sources are presented in Methods. \label{tab:R2}}
\end{table}

The center column shows the PDFs of the shape factor, $a_{r}-1$ (symbols).
The theoretical predictions per Eq. (\ref{eq:rhox}) are shown in
dashed, and exhibit excellent agreement with data. They are generated
per Eq. (\ref{eq:rhox}), with the single input parameter, $\beta$,
extracted from the analysis of $|{\bf E}|^{2}$ distribution. 

Lastly in the right column, the PDFs for $a_{r}-1$ are normalized
with $\left<a_{r}\right>-1$, where $\left<\cdot\right>$ denotes
a mean, e.g.,

\begin{equation}
\left<x\right>=\int_{0}^{\infty}x\rho_{X}(x)\mathrm{d}x.\label{eq:xm}
\end{equation}
Both data and theoretical predictions are normalized following this
practice. The dot-dashed are best fits using a $k$-Gamma distribution
defined as
\begin{equation}
\rho_{{\rm kG}}(x_{1};k)=\frac{k^{k}}{\Gamma(k)}x_{1}^{k-1}\exp(-kx_{1}),\label{eq:rhokG}
\end{equation}
where $\Gamma(k)$ is the Gamma function, and $k$ is the single fitting
parameter. The agreement is evident, and the $k$-values are found
to vary between 2 and 3. 

\begin{figure}
\centering\includegraphics[width=1\textwidth]{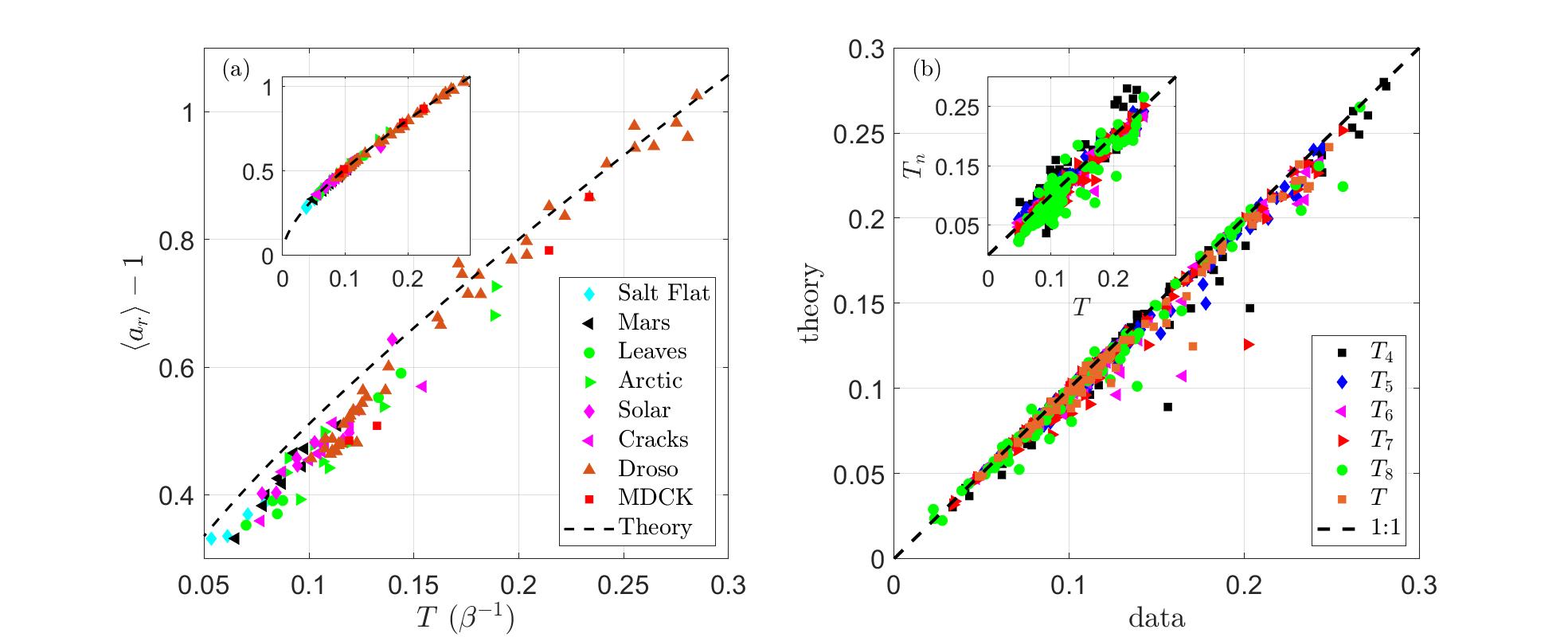}

\caption{(a) Overall correlation between $a_{r}$ and $T$ ( $\beta^{-1}$)
for all 99 data sets (symbols); the theoretical prediction (dashed)
is generated per (\ref{eq:rhox}). The inset alternatively presents
${\bf E}$ from a moment-based calculation which further improves
agreement. (b) A comparison between data and predicted temperature,
$T_{n}$ and $T$ (``theory''). The inset shows that sub-ensemble
temperature $T_{n}$'s are quantitatively similar to tessellation
temperature, $T$. \label{fig:Tx-All}}
\end{figure}

Overall corroboration between theory and data is quantified by the
coefficient of determination, $\mathrm{R}^{2}$, and are listed in
Table \ref{tab:R2} for all cases (see Methods for definition). The
values are uniformly close to $1$ (``perfect agreement'') with
minimal variations from case to case. Validity of the theoretical
prediction (\ref{eq:rhox}) is also attested by Fig. \ref{fig:Tx-All}a.
Here we define a pseudo temperature $T$ as the inverse of $\beta$,
namely, $T=\beta^{-1}$. To compare with data, a theoretical prediction
is generated by using (\ref{eq:rhox}) in the integration (\ref{eq:xm}).

The above results validate that the E\textsuperscript{2} does follow
an exponential, Boltzmann-like distribution. The universality of this
distribution in all data sets, according to our theory, necessarily
leads to a universality of $k$-Gamma distributions for the aspect
ratio. That is, the validity extends beyond the confluent tissues
studied in \cite{Atia2018}, and to all systems we analyzed. As a
corollary, (\ref{eq:rhox}) provides a fundamental solution for the
aspect ratio, of which the $k$-Gamma distributions (\ref{eq:rhokG})
are convenient approximations. This solution does not normalize to
a single curve with a single $k$ value if fitted with Gamma distributions.
In fact, it predicts a positive correlation between $\beta$ and $k$
(Fig. S8). Qualitatively, this means that PDFs of aspect ratio ($a_{r}$,
or equivalently, $x$) with wider spread and greater mean (corresponding
to lower values of $\beta$) present themselves relatively to the
left after normalization (corresponding to lower values of $k$ and
noting that maxima occur at $1-1/k$ per (\ref{eq:rhokG})). This
trend is fully corroborated with data from our own work (Fig. \ref{fig:Master},
and Fig. S2 in SI, center and right columns) and Atia et al. \cite{Atia2018}
(Fig. 3 therein), as well as predictions from a self-propelled Voronoi
model in the supplemental information of the latter. In summary, the
variability in $k$ arises from the variability in $\beta$.

We remark that the agreement between our theoretical prediction and
data can be even better if we use a variation of the deformation tensor
computed as the square-root of the area moment tensor, ${\bf M}$,
via Eq. (\ref{eq:EM}) in the Methods section. This is not surprising,
as now both ${\bf E}$ and $a_{r}$ share the same origin, the agreement
shown in Fig. \ref{fig:Tx-All}a inset is near perfect. The slight
differences between the two definitions are due to higher order effects
that we theoretically and numerically demonstrate in the SI. Here
and below we focus on using (\ref{eq:cE}) for its apparent analytical
simplicity.

\subsection*{E\textsuperscript{2} distribution is a $\chi^{2}$-distribution}

In the second part of this work, we demonstrate the origin of the
highly regular statistical distribution in E\textsuperscript{2}.
Figure \ref{fig:Tx-All}b shows results comparing the pseudo-temperature
computed from Eq. (\ref{eq:E2}) (denoted ``data'') with the theoretical
prediction we develop below (denoted ``theory''). Here the subscript
$n$ denotes a restriction to the sub-ensemble of $n$-gons, $T_{n}:=\left<|{\bf E}|^{2}\right>_{n}.$
Polygons other than $n=4$-$8$ are of statistically insignificant
occurrences and not included in the evaluation.

The key relationship we utilize is a quadratic form to compute $|{\bf E}|^{2}$
given vertex displacement, ${\bf v}_{j}$, 
\begin{equation}
|{\bf E}|^{2}={\bf \hat{v}}\cdot{\bf \hat{C}}{\bf \hat{v}}.\label{eq:uCu}
\end{equation}
Here ${\bf \hat{v}}$ is a concatenated vector in $\mathbb{R}^{2n}$,
\begin{equation}
{\bf \hat{v}}:=\left[\begin{array}{c}
\hat{v}_{1}\\
\vdots\\
\hat{v}_{2n}
\end{array}\right]=\left[\begin{array}{c}
{\bf v}_{1}\\
\vdots\\
{\bf v}_{n}
\end{array}\right].\label{eq:Concat}
\end{equation}
Other vectors such as ${\bf \hat{y}}$, ${\bf \hat{u}}$, and ${\bf \hat{e}}$
are similarly defined from their two-dimensional counterparts, and
all vectors and tensors in the $2n$-dimensional space are denoted
by a hat to differentiate from the planar quantities. ${\bf \hat{C}}\in\mathbb{R}_{{\rm sym}}^{2n\times2n}$
is a second-order tensor with block components ${\bf C}_{ij}$ given
in (\ref{eq:E2}). Not surprisingly, ${\bf \hat{C}}$ has 2 non-trivial
eigenvalues (SI), matching the degree of freedom of ${\bf E}$ (note
that ${\bf E}=(E_{11,}E_{12};E_{12},-E_{11})$ in a general component
form):
\[
{\bf \hat{P}}^{{\rm T}}{\bf \hat{C}}{\bf \hat{P}}={\rm diag}(2/n,2/n,0,...,0).
\]
The diagonalization above with the orthonormal tensor ${\bf \hat{P}}$
helps us express $|{\bf E}|^{2}$ in a particularly simple form, 
\begin{equation}
|{\bf E}|^{2}=\frac{2}{n}\left(\hat{w}_{1}^{2}+\hat{w}_{2}^{2}\right),\quad{\bf \hat{w}}:={\bf \hat{P}}^{{\rm T}}{\bf \hat{v}}.\label{eq:E2vtilde}
\end{equation}
We realize that $\hat{w}_{k}={\bf \hat{p}}_{k}\cdot{\bf \hat{v}}$,
${\bf \hat{p}}_{k}$ being an eigenvector of ${\bf \hat{C}}$. If
${\bf \hat{v}}$ is characterized by a covariance matrix, ${\bf \hat{\Sigma}}$,
then the variance of $\hat{w}_{k}$ is \cite{JohnsonWichern2007}
\[
{\rm Var}(\hat{w}_{k})={\bf \hat{p}}_{k}\cdot{\bf \hat{\Sigma}}{\bf \hat{p}}_{k}={\rm Tr}({\bf \hat{p}}_{k}\otimes{\bf \hat{p}}_{k}\cdot{\bf \hat{\Sigma}}),
\]
and $T_{n}$ is readily calculated as 
\begin{equation}
T_{n}=\left<|{\bf E}|^{2}\right>_{n}=\frac{2}{n}{\rm Var}(\hat{w}_{1}+\hat{w}_{2})={\rm Tr}({\bf \hat{C}}{\bf \hat{\Sigma}}).\label{eq:CSig}
\end{equation}
Note we have used ${\bf \hat{C}}=\frac{2}{n}({\bf \hat{p}}_{1}\otimes{\bf \hat{p}}_{1}+{\bf \hat{p}}_{2}\otimes{\bf \hat{p}}_{2}).$
Here and after and as a good approximation, we assume ${\bf \hat{v}}$
has a zero mean. Equation (\ref{eq:CSig}) is a \emph{precise} expression
to compute $T_{n}$ given ${\bf \hat{\Sigma}}$, and is used to generate
the theoretical predictions in Fig. \ref{fig:Tx-All}b, main panel.
The tessellation average $T$ can be computed by taking the weighted
sum of $T_{n},$ namely, $T=\sum_{n}(N_{n}T_{n})/\sum_{n}N_{n}$,
where $N_{n}$ is the number of $n$-gons. On the other hand, sub-ensemble
temperatures are typically quantitatively similar to the tessellation
temperature, as shown in Fig. \ref{fig:Tx-All}b inset.

If we further assume that $\hat{w}_{1,2}$ follow identical normal
distributions, immediately we have
\begin{equation}
\rho_{{\rm E}}(|{\bf E}|^{2})=\frac{1}{T_{n}}\exp\left(-\frac{|{\bf E}|^{2}}{T_{n}}\right).\label{eq:chi-square}
\end{equation}
In other words, the exponential distribution arises actually as a
$\chi^{2}$-distribution with 2 degrees of freedom. On the other hand,
if the variances ${\rm Var}(\hat{w}_{1,2})$ are not identical but
quantitatively similar, which is true for all tessellations we study
(see Fig. S5), Eq. (\ref{eq:chi-square}) still holds to the leading
order. (This point is straightforward to prove via Taylor expansion
and not shown here for brevity.) Note that even in this situation,
per (\ref{eq:CSig}) the formula for $T_{n}$ is still accurate without
approximation. This provides an essential illustration of the origin
of the E\textsuperscript{2} distribution, and Eq. (\ref{eq:chi-square})
is a main result of the current work. It remains to be shown below
that $\hat{w}_{1,2}$ distributions are indeed approximately normal
and independent.

\subsection*{Asymptotic normality contributes to statistical regularity}

\begin{figure}
\centering\includegraphics[width=1\textwidth]{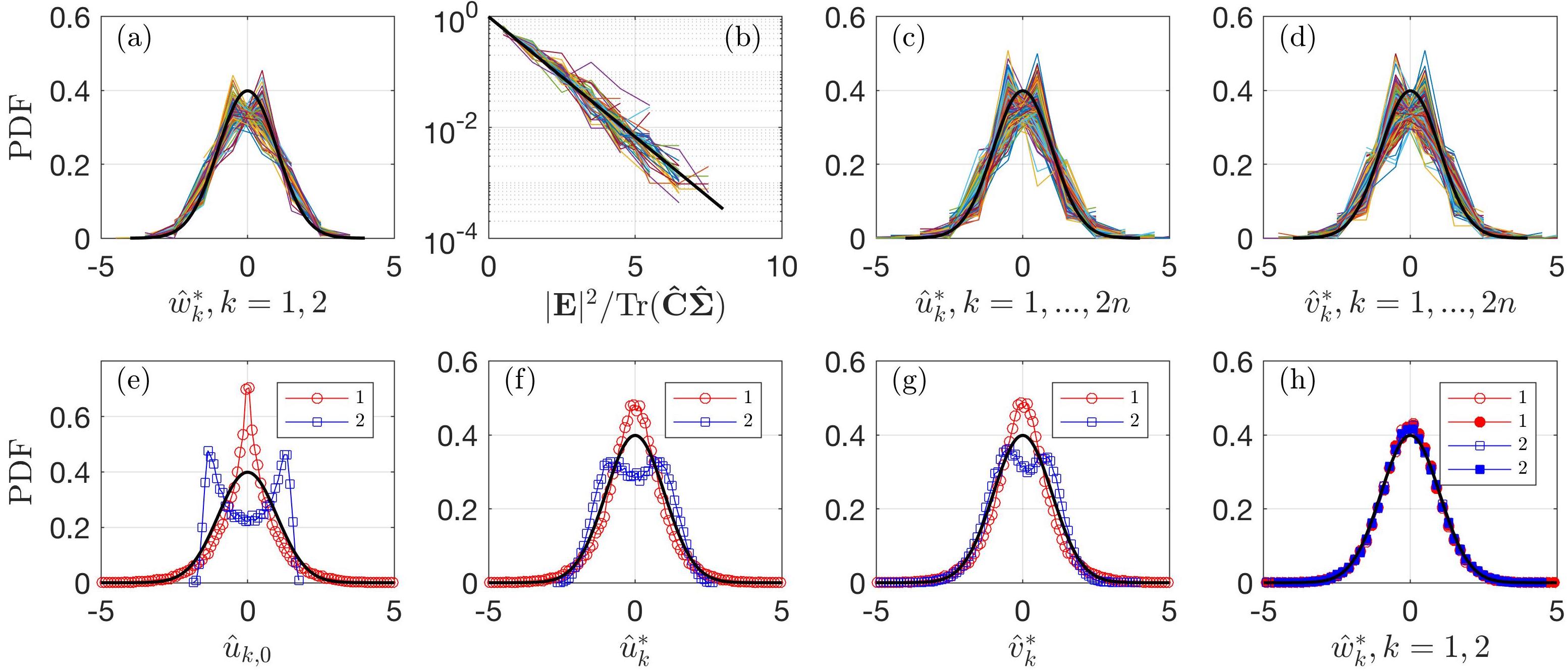}

\caption{Normality of key variables. Here the superscript $^{*}$ denotes normalization
by its own standard deviation so as to compare with $N(0,1)$, the
dark line in all panels except for (b) where it represents an exponential
function, $\exp(\cdot)$. (a) PDF for $\hat{w}{}_{k}$, $k=1,2$,
from all tessellations (198 profiles). (b) Normalized $|{\bf E}|^{2}$
follows a simple exponential distribution (99 profiles). (c) and (d)
demonstrate the normality of $\hat{u}_{k}$ and $\hat{v}_{k}$ ($k=1,...,2n$)
in the hexagon sub-ensemble ($n=6)$ , 1188 profiles each. (e) Initial
displacement distributions for two exemplary cases, 1 and 2. (f) and
(g) $\hat{u}_{k}$ and $\hat{v}_{k}$ asymptote toward normality.
(g) Normalities of $\hat{w}_{1}$ and $\hat{w}_{2}$ are well-established.
\label{fig:UND}}
\end{figure}

Fig. \ref{fig:UND}a presents $\hat{w}_{1,2}$ in the hexagon sub-ensemble
($n=6$) for all 99 tessellations, whereas the cases for $n=5$ and
7 are included in the SI. PDFs are all normalized for comparison with
standard Gaussian, $N(0,1)$ (dark solid lines). Although the PDFs
exhibit appreciable fluctuations due to the relatively small sample
size in the $n$-sub-ensemble, the approximate normalities are evident.
Quantitative similarities of $\hat{w}_{1,2}$ are demonstrated in
Fig. S5. In addition, $\hat{w}_{1,2}$ are indeed only weakly dependent,
as ${\rm Cov}(\hat{w}_{1},\hat{w}_{2})/{\rm Var}(\hat{w}_{1})=0.078\pm0.046$
for all cases, consistent with the anticipated 2 degrees of freedom.
All E\textsuperscript{2} distributions (normalized by the predicted
temperature $T_{n}$) are shown in Fig. \ref{fig:UND}b.

The apparent candidate to explain the resulting normality is the central
limit theorem in the generalized version for dependent and identical
random variables \cite{Lehmann1999}, noting that $\hat{w}_{k}$ derives
from $\hat{v}_{k}$ via a linear combination (\ref{eq:E2vtilde}).
It is peculiar to note that $\hat{u}_{k}$ and $\hat{v}_{k}$ themselves
also demonstrate approximate normality, shown in Fig. \ref{fig:UND}c
and d. The normality in $\hat{u}_{k}$ is again be explained by the
central limit theorem. We can write ${\bf \hat{u}}$, the concatenated
vector for ${\bf u}_{j}$ 's as (SI) 
\[
{\bf \hat{u}}={\bf {\bf \hat{R}}}({\bf \hat{y}}-\left<{\bf \hat{y}}\right>),\quad{\bf \hat{R}}:={\bf \hat{I}}-\frac{1}{n}{\bf \hat{e}}\otimes{\bf \hat{e}}.
\]
In the absence of apparent anisotropy, components of ${\bf \hat{y}}-\left<{\bf \hat{y}}\right>$
are approximately dependent yet identical, satisfying precondition
of the theorem. Hence $\hat{u}_{k}$ asymptotes to normality. On the
other hand, from Eq. (\ref{eq:vi}) we have ${\bf \hat{v}}={\bf \hat{u}}/r_{0}$
and $r_{0}=({\bf \hat{y}}\cdot{\bf \hat{e}})/n$. The normality in
${\bf \hat{v}}$ is difficult to theoretically prove. However, it
is reasonable to speculate the loss of the apparent scale would create
similarity to preserve or even enhance normality - see also Figs \ref{fig:UND}f,
g below. In addition, it is extensively confirmed by the data as shown
in Fig. \ref{fig:UND}d.

The asymptotic normality can be better illustrated via a simple Monte
Carlo simulation following the schematic in Fig. \ref{fig:Schematic}k,
where we temporarily restrict to an isolated hexagon, and displacements
$\hat{u}_{k,0}$ ($k=1,2,...,2n$) are prescribed according to independent,
identical distributions as shown in Fig. \ref{fig:UND}e (Methods,
Eqs. (\ref{eq:ui0}, \ref{eq:di})). Two representative cases are
examined, the first with a steeper than Gaussian initial descent,
and the second non-monotonic. In Figs. \ref{fig:UND}f and g, both
$\hat{u}_{k}$ and $\hat{v}_{k}$ already demonstrate trending toward
normality, although some differences from $N(0,1)$ are still visible.
Note only an arbitrary index $k$ is shown as these distributions
are expected to be identical. Subsequently, in Fig. \ref{fig:UND}h
the normality of $\hat{w}_{1,2}$ are well established. $|{\bf E}|^{2}$
distribution quantitatively follows our theoretical prediction and
is not shown for brevity. Although only two exemplary tests are presented,
repeated simulations reveal the same asymptotic trend to normality
and the quantitative relationships (\ref{eq:CSig}, \ref{eq:chi-square})
always hold.

In a summary, the above exercises demonstrate that asymptotic normality
is prevalent in planar tessellations, as key variables derive from
linear combinations of statistically similar components. As a result
E\textsuperscript{2} distributions become highly regular due to combined
normality and its low-dimensionality.

\subsection*{Physical Implications}

The physical meaning of ${\bf E}$ is self-evident: it represents
deformation, and hence is typically associated with energy in one
form or another. In the wide range of phenomena we studied, the constitutive
relations come in different forms (some are yet unknown). However,
some usual possibilities can be contemplated. If the energy is bulk-elastic
in nature, then any physically reasonable elastic model of a polygon,
valid in the small-to-moderate deformation regime, \emph{must} follow
the form \cite{Gurtin2010}
\begin{equation}
\Delta\Psi=\mu|{\bf E}|^{2},\label{eq:Constitutive-1}
\end{equation}
where $\mu$ is the first Lam\'{e} constant, whereas the second constant
is not needed as ${\rm Tr}{\bf E}=0$ (SI). On the other hand, if
energy is associated with edge lengths or perimeters, such as in the
case of models for 2D confluent tissues \cite{Bi2015,KimHilgenfeldtSM2015,Manning_2010},
Eq. (\ref{eq:Constitutive-1}) is still a formally valid approximation,
as the change in perimeter is also proportional to $|{\bf E}|^{2}$
to leading-order approximation (SI). Last but not least, in the Quantizer
problem \cite{Du2005Gersho,HainKlattTurk2020,KlattNatCom2019} the
cell-wise energy functional is the moment of inertia, which is
\begin{equation}
{\rm Tr}{\bf M}=2m_{0}(1+|{\bf E}|^{2})\label{eq:TrM}
\end{equation}
in both two and three dimensions (SI), and $2m_{0}$ is the moment
of inertia of the regular reference polygon, $\mathcal{P}_{R}$. Thus
its distribution can be computed via knowing both the area and E\textsuperscript{2}
distributions. These examples of constitutive relations cover a reasonably
wide range of physical systems.

Above we have taken the liberty in naming a pseudo-temperature, $T$
(or $T_{n}$ for the sub-ensembles). Indeed, such definition is both
tempting and appropriate in the presence of a Boltzmann-like distribution.
The tests by Dean and Lef\`evre \cite{DeanLefevre2003} and McNamara
\emph{et al. }\cite{McNamara2009} become trivial: the ratio of two
overlapping exponential distributions will necessarily give another
exponential distribution. We therefore name this pseudo-temperature
the ``E\textsuperscript{2} temperature'', and propose it as a candidate
for a thermodynamically meaningful temperature owing to its consistent
regularity and connection with physical quantities. This temperature
quantifies the overall deformation, and possibly also system energy.
To further explore a thermodynamic framework would require system-specific
physical principles, e.g., energy minimization, which we shall explore
in future work.

\vspace{1cm}

We have thus demonstrated three main points in this work: \emph{i})
An exponential distribution in E\textsuperscript{2} leads to a $k-$Gamma
distribution in the aspect ratio. In fact, $k$-Gamma distributions
are mere approximations to the more basic solution we develop. \emph{ii})
E\textsuperscript{2} distribution is a $\chi^{2}$-distribution with
two degrees of freedom, arising from combined effects of asymptotic
normality and the small dimensionality of ${\bf E}$, which is analogous
to the small dimensionality of the volume function in granular assembly
\cite{BlumenfeldEdwards2003}. We have developed a formula to compute
E\textsuperscript{2}\emph{ }from vertex statistics. \emph{iii}) E\textsuperscript{2}
and aspect ratio distributions as well as normality in displacements
are true universal features as we have shown via both a large collection
of data and theoretical derivations illustrating their mechanistic
origins. The strong regularity in E\textsuperscript{2} and vertex
displacements are ``hidden patterns'' revealed by this work. The
mean-field strain tensor, with its clear physical and geometric meaning,
is an ideal quantity connecting the conservation principles, the energy
(or pseudo-energy) landscape, and the geometric distributions. It
is a powerful quantifier to describe polygonal networks randomized
by active agitations, structural defects, and noises, among others.
Analysis may also be extended to polytopes in three and higher dimensions.

\subsection*{Author contributions}

HL, LL, and KDI designed the research; HL, LL, RL, and SM developed
the theory; RL, ZZ, and CI analyzed images; HL, RL, and SM analyzed
data; CI and ZZ performed experiments; HL, LL, and KDI wrote the paper. 

\subsection*{Acknowledgement}

The authors are grateful to Dr. Yuanwang Pan for providing fixed wing
disc images in the Droso group. The authors acknowledge helpful discussions
with Dr. Troy Shinbrot, and funding support from NIH R21 CA220202-02
(PI: HL); NSF CMMI 1351561 and DMS 1410273 (PI: LL); and NIH R35 GM131748
(PI: KDI). 

\subsection*{Competing interests}

The authors declare no competing interests.

\subsection*{Data availability}

The datasets generated during and/or analysed during the current study
are available from the corresponding authors on reasonable request.

\subsection*{Methods}

\subsubsection*{Data Collection}

Among the data groups listed in Table 1, the last 3 (Droso Fix, Droso
Live, and MDCK) are generated from this work, whereas other data types
are collected from the public domain. They are briefly described below,
and the specific images analyzed are identified in the references
where possible.

\paragraph*{Salt Flat of Uyuni (Salt Flat)}

All images of the Salar de Uyuni (Bolivia) are from online, or captured
from still frames of online videos. Credits are given to identifiable
author IDs, and time stamps in videos are provided \cite{SaltFlat4,SaltFlat2,SaltFlat1,SaltFlat6,SaltFlat5,SaltFlat3,SaltFlat7}.

\paragraph*{Landforms on Mars (Mars)}

All photos come from the High Resolution Imaging Science Experiment
(HiRISE) on board the Mars Reconnaissance Orbiter and are produced
by NASA, JPL-CalTech and University of Arizona \cite{Levy2009,ElMaarry2010,Mars2,Mars3,Mars5,Soare2019}.
The mechanisms of geological pattern formation on Mars are still the
subject of active studies, and theories include desiccation \cite{ElMaarry2010},
thermal contraction \cite{Levy2009,Mars2,Mars3}, and ice sublimation
\cite{Soare2019}. The image from \cite{Mars5} likely indicates \textcolor{black}{ridges
of sand dunes.}

\paragraph*{Veins on Leaves (Leaves)}

All images are from online where proper credits are given to website,
author, or author ID whichever is identifiable \cite{Leaf2,Leaf3,Leaf4,Leaf1,Leaf5,Leaf6}.
Species are not identified in photos except for \cite{Leaf2}, which
shows \emph{Ficus lyrata} (Fiddle-Leaf Fig).

\paragraph*{Landforms in the Arctic (Arctic)}

Polygonal landforms in or near the Arctic are mostly ice wedges \cite{Ice6,Maison2018,Liljedahl2016,Ice10,Ice4,Ice7,Worsley2014}
or tundra \cite{Aleina2014,Ice11}, whereas patterns in the latter
typically corroborate with locations of ice wedges, too.

\paragraph*{Supergranulation on Solar Surface (Solar)}

Supergranulation patterns on the solar surface from observations \cite{Berrilli1999,Chatterjee2017,DeRosa2004,Hagenaar1997,Hirzberger2008,Schrijver1997}.

\paragraph*{Cracked dry earth (Cracks)}

Land cracks, mostly probably formed due to desiccation. Images are
collected from the internet \cite{Crack5,Crack2,Crack6,Crack8,Crack11,Crack1,Crack10,Wang2013,Crack9}.

\paragraph*{Drosophila wing disc, fixed (Droso)}

Drosophila were cultured at 25°C. To obtain fixed wing discs at different
stages, eggs were laid for 2 to 4h, and larvae were dissected at 72,
84, 96, 108 and 120h after egg laying (AEL). Dissected wing discs
were fixed in 4\% paraformaldehyde for 15 min at room temperature.
Staining of fixed wing discs was performed essentially as described
in \cite{Rauskolb2018} using rat anti-E-cad (1:400 DCAD2; DSHB) and
anti-rat Alexa Fluor 647 (Jackson ImmunoResearch, 712-605-153). Images
were captured on a Leica SP8 confocal microscope. To compensate for
aberrations due to the curvature of wing disc and signals from the
peripodial epithelium, we used the Matlab toolbox ImSAnE \cite{Heemskerk2015}
to detect and isolate a slice of the wing disc epithelium surrounding
the adherens junctions, which was then projected into a flat plane,
as described previously \cite{Pan2016}. 

\paragraph*{Drosophila wing disc, live (Droso)}

For live imaging of cultured wing discs, larvae expressing GFP-labelled
E-cadherin from a Ubi-Ecad:GFP transgene were dissected at 96h AEL,
and then cultured based on the procedure of Dye \emph{et al.} \cite{Dye2017}.
Live wing discs were imaged using a Perkin Elmer Ultraview spinning
disc confocal microscope every 8 mins for 12 hours.

\paragraph*{Plated MDCKIIG cells (MDCK)}

MDCKIIG (a gift from W. James Nelson, Stanford University) cells were
cultured in low-glucose Dubecco\textquoteright s modified Eagle\textquoteright s
medium (DMEM) (Life Technologies) supplemented with 10\% fetal bovine
serum (FBS) and antibiotic-antimycotic. Cells were used at low passage
number, checked regularly for contamination by cell morphology and
mycoplasma testing. Cells were plated at different densities (1.5,
3, 4.5, 6, and 7.5$\times10^{4}$ cells/cm$^{2}$) on coverslips coated
with 0.6 mg/ml of collagen for 15 min at room temperature and washed
with PBS. After 48 hours, cells were fixed with 4\% paraformaldehyde
in PBS$++$ (phosphate- buffered saline supplemented with 100 mM MgCl$_{2}$
and 50 mM CaCl$_{2}$) for 10 min at room temperature. Immunostaining
was performed as in Ibar \emph{et al.} using mouse anti-ZO1 (1:1000,
Life Technologies \#33-9100) and anti-mouse Alexa Fluor 647 (Jackson
ImmunoResearch) \cite{Ibar2018}. Images were acquired using LAS X
software on a Leica TCS SP8 confocal microscope system using a HC
PL APO 63$\times$/1.40 objective. 

\subsubsection*{Image and data analysis}

Fluorescent images are analyzed using Tissue Analyzer, a plug-in of
ImageJ (version 1.52j), from which the cells are sectioned and cell
centroids, edges, and vertices are identified. Post-processing is
then performed with MATLAB. For each cell, the second area moment
tensor, defined with respect to the cell centroid ${\bf c}$, is 
\begin{equation}
{\bf M}=\int_{\mathcal{P}}({\bf y}-{\bf c})\otimes({\bf y}-{\bf c})\mathrm{d}A,\label{eq:areamoment}
\end{equation}
where the integration is over the polygon (cell) $\mathcal{P}$. Note
that here we ignore the curvature of cell edges and assume (by approximation)
that they are straight lines connecting vertices. Standard and exact
formulae are available for polygons which we use to compute the components
of ${\bf M}$ with only the coordinates of the vertices, ${\bf y}_{j}$'s.
The aspect ratio $a_{r}$ is 
\[
a_{r}=\sqrt{\frac{\max(\lambda_{1},\lambda_{2})}{\min(\lambda_{1},\lambda_{2})}},
\]
where $\lambda_{1}$ and $\lambda_{2}$ are eigenvalues of ${\bf M}$.

As an alternative approach to calculate ${\bf E}$, we could bypass
${\bf F}$ and make use of the moment. We denote this definition ${\bf E}_{{\rm M}}$,
\begin{equation}
{\bf E}_{{\rm M}}=\left(\frac{{\bf M}}{\sqrt{\mathrm{det}{\bf M}}}\right)^{\frac{1}{2}}-{\bf I}.\label{eq:EM}
\end{equation}
In the SI we demonstrate that to the leading order the two definitions
are approximately equal. We note that while (\ref{eq:EM}) is an area-based
calculation, (\ref{eq:cE}) in the proper text is vertex-based.

The coefficient of determination, $\mathrm{R}^{2}$, follows the standard
definition, 
\[
\mathrm{R}^{2}=1-\frac{\mathrm{Var}({\bf f}-{\bf f}')}{\mathrm{Var}({\bf f})}.
\]
Here '$\mathrm{Var}$' denotes variance, ${\bf f}$ is the data presented
in array form, and ${\bf f}'$ is the corresponding array generated
via fitting (such as for $|{\bf E}|^{2}$) or a theoretical prediction
(such as for $a_{r}$).

\subsubsection*{Monte Carlo simulation}

Results shown in Fig. \ref{fig:UND}e-h are generated via a simple
Monte Carlo simulation. We generate deformation from regular hexagons
using Eq. (\ref{eq:uj}). The initial displacements ${\bf u}_{i}^{0}$
follow independent and identical distributions 
\begin{equation}
{\bf u}_{i,0}=(d_{i}\cos\theta_{i},d_{i}\sin\theta_{i}),\label{eq:ui0}
\end{equation}
where $\theta_{j}$ is uniformly distributed between $[0,\:2\pi]$,
and $d_{i}$ follows 
\begin{equation}
d_{i}\sim\beta pd_{i}^{p-1}\exp(-\beta d_{i}^{p}).\label{eq:di}
\end{equation}
We set $\beta$ to ensure the variances of the components are 1. The
exponent $p=4/3$ gives test case 1 (red) shown in Fig. \ref{fig:UND}e,
whereas $p=10$ gives test case 2 (blue), which has a non-monotonic
distribution of $d_{i}$, and hence $\hat{u}_{k,0}$. These random
initial displacements lead to a centroid-translation (SI), 
\[
{\bf c}=\frac{1}{n}\sum_{i=1}^{n}{\bf u}_{i,0}.
\]
Center-correction of ${\bf u}_{i,0}$ provides ${\bf u}_{i}$ and
${\bf y}_{i}$ in (\ref{eq:uj}). Once ${\bf y}_{i}$'s are generated,
other quantities such as $\hat{v}_{i}$, $\hat{w}_{i}$ and $|{\bf E}|^{2}$
are computed according to formula provided in the proper text.

\newpage

\bibliographystyle{plain}
\bibliography{E2paper_Nov2020}

\end{document}